\documentclass[]{spie}

\usepackage[]{graphicx,color}
\usepackage[]{amsmath}
\usepackage[]{amssymb}

\newcommand{\farcs}{\hbox{$.\!\!^{\prime\prime}$}}
\newcommand{\sun}{\hbox{$\odot$}}

\title{High-contrast imaging in the Hyades with snapshot LOCI}

\author{
Katie M. Morzinski\authorinfo{\supit{*}Sagan Fellow.  Contact: ktmorz@arizona.edu}\supit{*a},
Bruce A. Macintosh\supit{b},
Laird M. Close\supit{a},
Christian Marois\supit{c},
Quinn Konopacky\supit{d},
and
Jenny Patience\supit{e}
\skiplinehalf
\supit{a} Steward Observatory, University of Arizona, Tucson, AZ;
\skiplinehalf
\supit{b} Lawrence Livermore National Laboratory, Livermore, CA;
\skiplinehalf
\supit{c} Herzberg Institute of Astrophysics, NRC-CNRC, Victoria, BC;
\skiplinehalf
\supit{d} Dunlap Institute for Astronomy and Astrophysics, Toronto, ON;
\skiplinehalf
\supit{e} School of Earth \& Space Exploration, Arizona State University, Tempe, AZ
}

\begin{document} 
\maketitle

\begin{abstract}
To image faint substellar companions obscured by the stellar halo and speckles, scattered light from the bright primary star must be removed in hardware or software.
We apply the ``locally-optimized combination of images'' (LOCI) algorithm to 1-minute Keck Observatory snapshots of GKM dwarfs in the Hyades using source diversity to determine the most likely PSF.
We obtain a mean contrast of $10^{-2}$ at 0\farcs01, $10^{-4}$ at $<$1'', and $10^{-5}$ at 5''.
New brown dwarf and low-mass stellar companions to Hyades primaries are found in a third of the 84 targeted systems.
This campaign shows the efficacy of LOCI on snapshot imaging as well as on bright wide binaries with off-axis LOCI, reaching contrasts sufficient for imaging 625-Myr late-L/early-T dwarfs purely in post-processing.
\end{abstract}

\section{INTRODUCTION}
Brown dwarfs have degenerate cores, convective interiors, and clear to cloudy atmospheres---intermediate properties between stars and planets.
Their distribution is also intermediate: despite being easier to detect, brown dwarf counts\cite{dwarfarchives.org} are already being overtaken by exoplanets\cite{exoplanet.eu}.
In order to better understand these ``failed stars,'' we seek to discover brown dwarfs in a known environment that can act as a controlled laboratory.
Toward that end, we undertake here a survey for brown dwarfs as companions to sun-like stars in the Hyades star cluster.

The Hyades is a nearby ($\sim$46 pc) open star cluster with about 400 members\cite{roeser2011}.
Because its age and metallicity are known, stellar models can accurately determine the mass of a star given its luminosity.
Because it is nearby and has a high proper motion ($\sim$0\farcs11/yr)\cite{perryman1998}, direct-imaging surveys are fruitful in that companionship can be determined based on co-moving proper motion.
Thus, we select 84 GKM Hyads for a direct-imaging search for low-mass companions, with sensitivity down to the L-T transition.

In this survey, we employ Keck Observatory adaptive optics (AO) with the NIRC2 camera in natural guide star (NGS) mode.
Images are acquired using ``snapshots'' of shallower (0.018~s $\times$ 500 co-adds) and deeper (5~s $\times$ 6 co-adds) exposures taken over the course of 3 nights in the first epoch (2005/08/23, 2006/01/15, and 2006/12/10), spending only 5--10 minutes total observing time (including overhead) per target.
A second epoch of observations are undertaken in 2009--2012, with Lick/IRCAL and MMT/ARIES observations as well as Keck/NIRC2.

The target stars are PSF-subtracted to reveal faint companions, using the locally-optimized combination of images\cite{loci} (LOCI) algorithm.
The LOCI algorithm determines the best estimate of the point-spread function (PSF) for a particular image, using a least-squares criterion.
In this work we employ LOCI in two unique ways:
1) Snapshot LOCI: Short-exposure snapshots are taken with a fixed position angle, and the other target stars serve as the reference library for the image of interest.
2) Off-axis LOCI: We also find off-axis PSFs using LOCI on bright wide binaries, enabling us to search the secondary star for close faint companions.

In this paper we describe implementation of the LOCI algorithm with fixed-position-angle AO snapshots,
and present our preliminary results of the search for low-mass companions in the Hyades.

\section{Principles of snapshot LOCI}
The point-spread function is the response of the imaging system to a delta-function input.
According to linear systems theory, the output image is the sum of the input sources.
Therefore, if we can determine the impulse response, we can then subtract the PSF to reveal other sources in the image.

Various methods are used to determine the PSF.
Many rely on some type of image diversity, in which a variety of images are taken under sufficient diversity of circumstances to be able to isolate the impulse response.
For example, angular differential imaging (ADI, \textit{a.k.a.}\ roll subtraction) uses diversity of position angle
while spectral differential imaging (SDI) uses diversity of wavelength.

In this work we use source diversity, in which a library of reference stars is used to provide a sampling of point sources.
We term this ``snapshot LOCI'' because hundreds of images taken at a rate of 1 target per 5--10 minutes make up the reference library.

\subsection{LOCI introduction}
The LOCI algorithm is applied to some set of image-diversity observations.
Out of a library of on order 100 images, it builds up the best estimate of the PSF.
``Best'' here is defined as that which gives the minimum residual noise when differenced with the target image.
LOCI uses a least-squares approach to construct the best PSF.
The PSF is built in smaller partitions, then jigsawed back together.  Using localized regions allows for independence in the PSF determination across the image plane.  The local regions, however, must be significantly larger than a $\lambda/D$-sized FWHM; otherwise, a single speckle could be found in a small region that could subtract off the faint companion.

For ground-based AO observations, the Strehl must be moderately high ($H$-band Strehls in our Keck data are typically 30--50\%) such that quasi-static speckles dominate over random atmospheric speckles.
Because photometric imaging is cumulative, different combinations of speckles may exist at a given time.
Thus, the algorithm searches for weights that can be applied to each reference image so that the weighted sum of the reference images gives the best PSF in aggregate.
These weights are computed as follows.

\subsection{LOCI\cite{loci} formalism}
We desire to find the PSF $P$ that minimizes the residual error when subtracted from the target star $T$, by taking a local region of each $k^\text{th}$ image $I ^k$ out of a total of $\ell$ in the reference library and summing with weights $w ^k$:
\begin{equation}
P = \sum_{k=1}^{\ell} w ^k I ^k .
\end{equation}

The variance $\sigma$ that we minimize for each $i^\text{th}$ pixel out of $n$ total pixels in the local region is given by
\begin{equation}
\sigma ^2 = \sum_{i=1}^{n} (T_i-P_i)^2 .
\end{equation}
To find the minimum variance, we find the weights where the derivative is equal to zero:
\begin{eqnarray}
\dfrac{d}{dw} \left( \sigma ^2 \right)
  &=& \dfrac{d}{dw} \sum_{i=1}^{n}  \left(  T_i - P_i  \right)^2     \\
  &=&  \dfrac{d}{dw} \sum_{i=1}^{n}  \left(  T_i - \sum_{k=1}^{\ell} w ^k I_i ^k  \right)^2  \nonumber  \\
  &=&  2 \sum_{i=1}^{n}  \left(  T_i - \sum_{k=1}^{\ell} w ^k I_i ^k \right)   \left(  -\sum_{j=1}^{\ell} w ^j I_i ^j  \right)
  =   0     \nonumber   .
\end{eqnarray}
Setting the derivative to zero, the minimum is given by
\begin{equation}
   \sum_{k=1}^{\ell} w ^k   \left(  \sum_{k=1}^{\ell} \sum_{i=1}^{n}   I_i ^k  \right)   \left(  \sum_{j=1}^{\ell} \sum_{i=1}^{n}   I_i ^j  \right)   
   =  \left(  \sum_{i=1}^{n} T_i \right)  \left(  \sum_{j=1}^{\ell} \sum_{i=1}^{n}   I_i ^j  \right)   ,
\end{equation}
or
\begin{equation}
 w ^k    [ I_i ^k   I_i ^j ] = [ T_i   I_i ^j ] 
\end{equation}
with the summations dropped for simplicity.

Solving for the weights that minimize the variance, we have
\begin{equation} \label{eq:weights}
 w ^k  =  [ I_i ^k   I_i ^j ]^{-1}  [ T_i   I_i ^j ]    .
\end{equation}

\subsection{LOCI intuition}
The matrix $[ I_i ^k   I_i ^j ]$ in Eqtn.~\ref{eq:weights} is an autocorrelation of the reference library images, whose value at a given pixel is higher where one image is bright, and highest where both images are bright.  (Look ahead to Fig.~\ref{fig:autocorrelation} for a visualization of this matrix.)
The matrix $[ T_i   I_i ^j ]$ is a cross-correlation of the target with the reference library, and its value is higher where both images are bright.

The LOCI algorithm is essentially a filter for image correlation.
The autocorrelation matrix $[ I_i ^k   I_i ^j ]$ searches for speckles occurring throughout all the images.
The cross-correlation matrix $[ T_i   I_i ^j ]$ searches for speckles that occur in the reference library and the target.
Finally, PSF subtraction filters out the systematic diffraction features, leaving behind uncorrelated any signal such as a planet or brown dwarf companion.

We invert the autocorrelation matrix, take its product with the cross-correlation, and sum over the pixels in the local region.
Then the PSF is built by summing over all the images in the library, weighted by the coefficients $w ^k$ that LOCI determines.
Finally, the local regions are stitched or jigsawed back together to give the best PSF across the image as a whole.

\section{Snapshot LOCI in practice}

\subsection{Compiling the reference library}
A unique reference library is compiled for each target star.
All images taken with NIRC2 for this survey are available to the reference library --- the observation date is not taken as a constraint.
Images must meet all of the following criteria to be included in the reference library:

\begin{enumerate}
\item For a given FITS file, all other images of that same star are excluded.  This provides for source diversity, ensuring that a brown dwarf will not self-subtract.
\item Close binaries are excluded.  All images contributing to the PSF must represent the impulse response to a delta function, so only single stars are allowed in the reference library.  (Close is a relative term but if there is a bright binary, the secondary must be on order 2'' away, such that it is beyond the majority of the quasistatic speckles of the primary.)
\item All images in the reference library for a given target must be through the same spectral filter as the target.  This is because speckles are a product of diffraction, and diffraction varies with wavelength.
\item Finally, we did impose the requirement that the reference library images be at the same approximate exposure time, so that there is a separate reference library for unsaturated (shallow) and saturated (deep) images.  However, we later find this criterion to be unnecessary, as we demonstrate in \S\ref{sec:locifluxnorm} that LOCI automatically accounts for varying flux levels.  Therefore, this criterion may be dropped provided the reference images can still be registered to sub-pixel accuracy (see \S\ref{sec:reflib}).
\end{enumerate}

\subsection{Preparing the reference library}   \label{sec:reflib}
The LOCI algorithm minimizes the least squares residuals for each pixel in the image.
Therefore, it is crucial that the PSFs in the reference library are registered to a common stellar center with sub-pixel precision.
We register the saturated and unsaturated images separately.

\begin{figure}[htbp]
  \begin{center}
      \includegraphics[height=0.75\linewidth,angle=90]{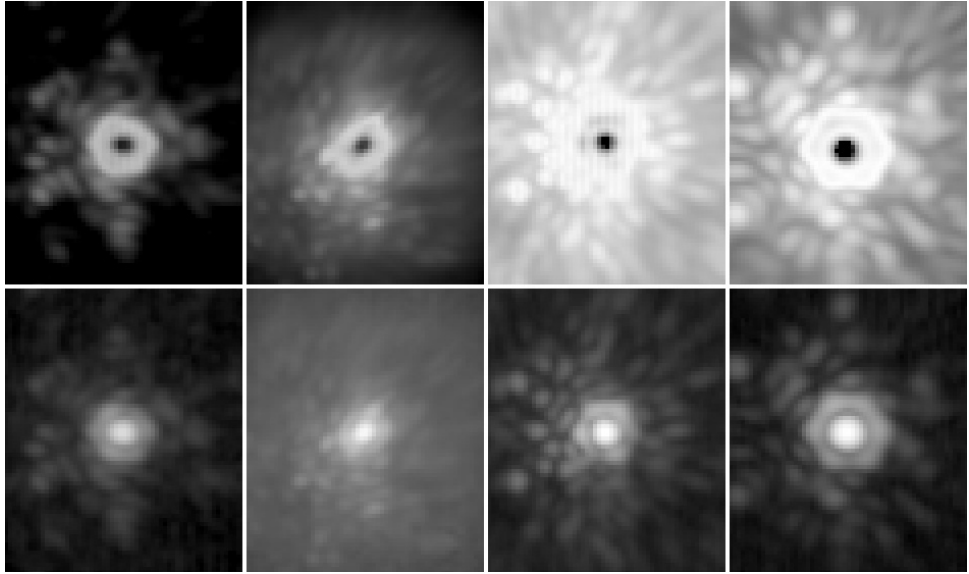}
  \end{center}
      \caption{
  	Fiducial images used for registration,
	which are selected for having among the highest Strehls and for being single point-source stars.
	Left to right: CH$_4$-short, J, H, K-prime.
	Top to bottom: Saturated and unsaturated.
	  } \label{fig:fiducials}
\end{figure}

Registration is a three-step process:
\begin{enumerate}
\item First, the stars are shifted to their approximate center, as determined by the location of maximum flux for short-exposure unsaturated images, or the minimum flux in the central saturated region of long-exposure images (see the saturated images in Fig.~\ref{fig:fiducials} for how saturation on NIRC2 results in an inverted peak).
\item Next, the image is autocorrelated with itself by rotating it by 180$^\circ$ and finding the point where the correlation is the maximum. This finds the center of the image to within one pixel.
If the camera position angle is rotated differently for the different stars, the images are next rotated so that the pupils are at the same angle and the six hexagonal diffraction spikes line up.
\item Finally, the sub pixel registration is done by choosing one of the images as a fiducial and then guessing and checking by shifting and subtracting the other images from the fiducial to find the best fit.  Figure~\ref{fig:fiducials} shows the fiducial images chosen for registration.
\end{enumerate}

\subsection{Reference library intuition}
\begin{figure}[htbp]
\centering
      \includegraphics[width=0.7\linewidth]{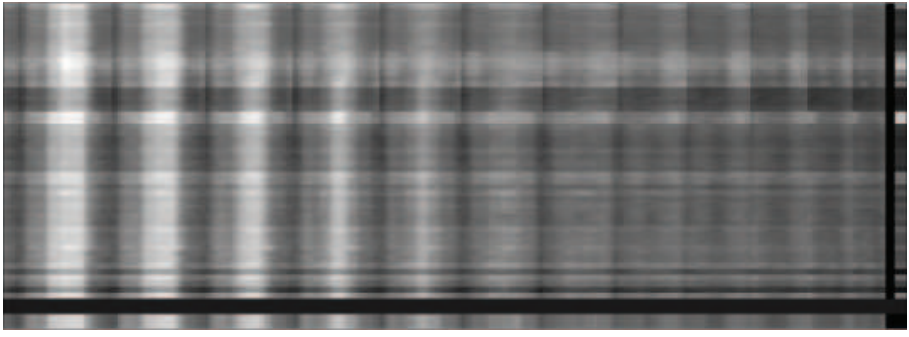}
      \caption{
  	Selection from a reference library.
	Each row represents a reference image.
	Each column represents a pixel in the image.
	The final row is the target image.
	The final column is the coefficients found by LOCI.
	  } \label{fig:reflib}
\end{figure}

Figure~\ref{fig:reflib} illustrates an example reference library for one localized region near the stellar core.  Each row represents a different image.  The bottom row is the target image; the others are reference images.  The images are a small localized region that is unwrapped into a vector for this display.  Each column is a given pixel on the detector.  
The alternating light and dark patches are speckles and Airy rings that line up on certain pixels.  Finally, the rightmost column represents the weights applied to each image (white being highest and black being lowest \textit{i.e.}\ most negative).

\subsection{Flux normalization in snapshot LOCI} \label{sec:locifluxnorm}
The value of the autocorrelation matrix $[ I_i ^k   I_i ^j ]$ is highest where the images are correlated in both structure and brightness.  Therefore, we investigate whether the flux of each star in the reference library should be normalized to improve LOCI performance.

We take a target image and construct its reference library (Fig.~\ref{fig:reflib}); this is Case 1.  We derive the best PSF with LOCI using the original reference library.  Then a second reference library is constructed in which we multiply the flux of reference image \#83 by a factor of 10; this is Case 2.  We derive the best PSF with LOCI using the second reference library.

\begin{figure}[htbp]
\centering
      \includegraphics[width=0.4\linewidth]{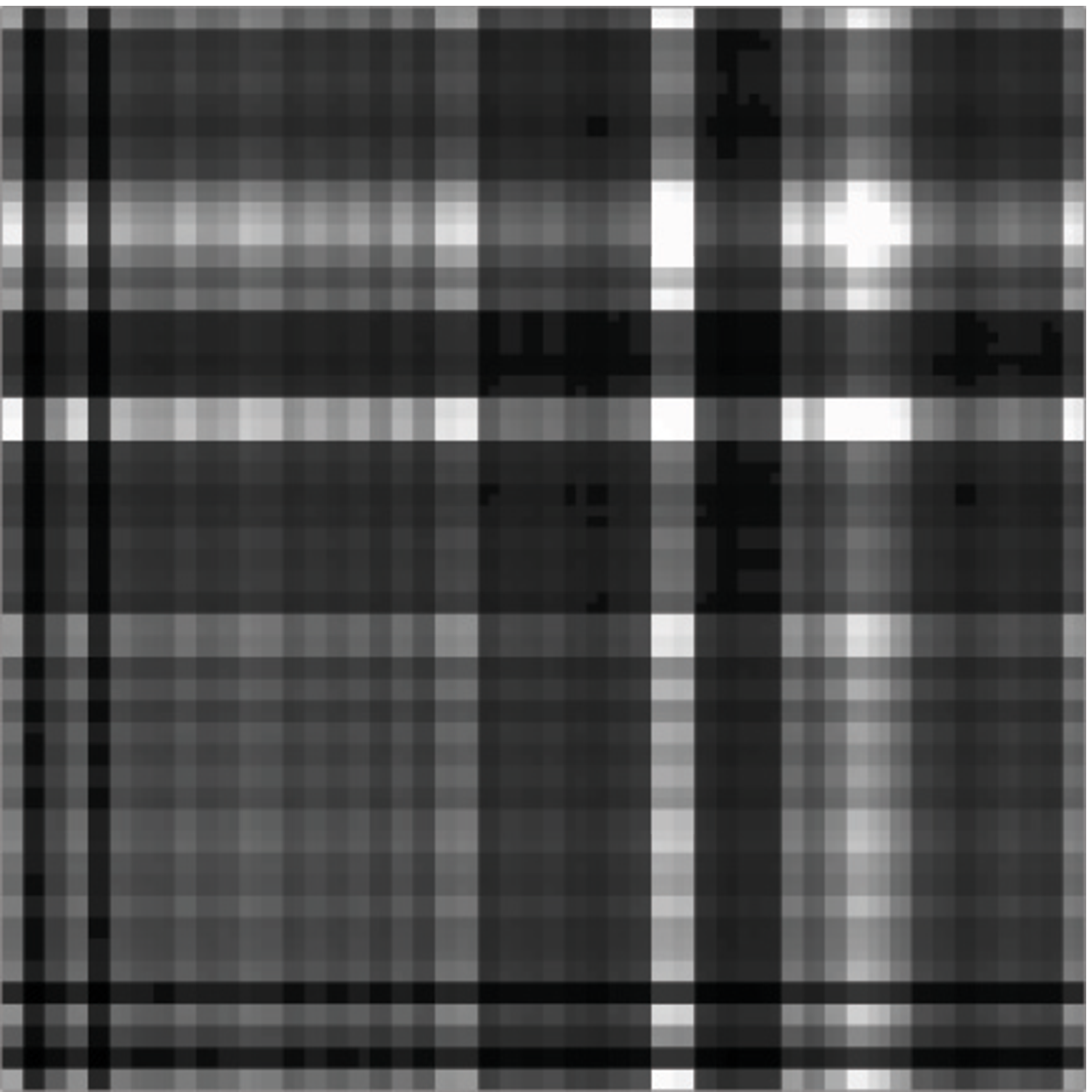}
      \includegraphics[width=0.4\linewidth]{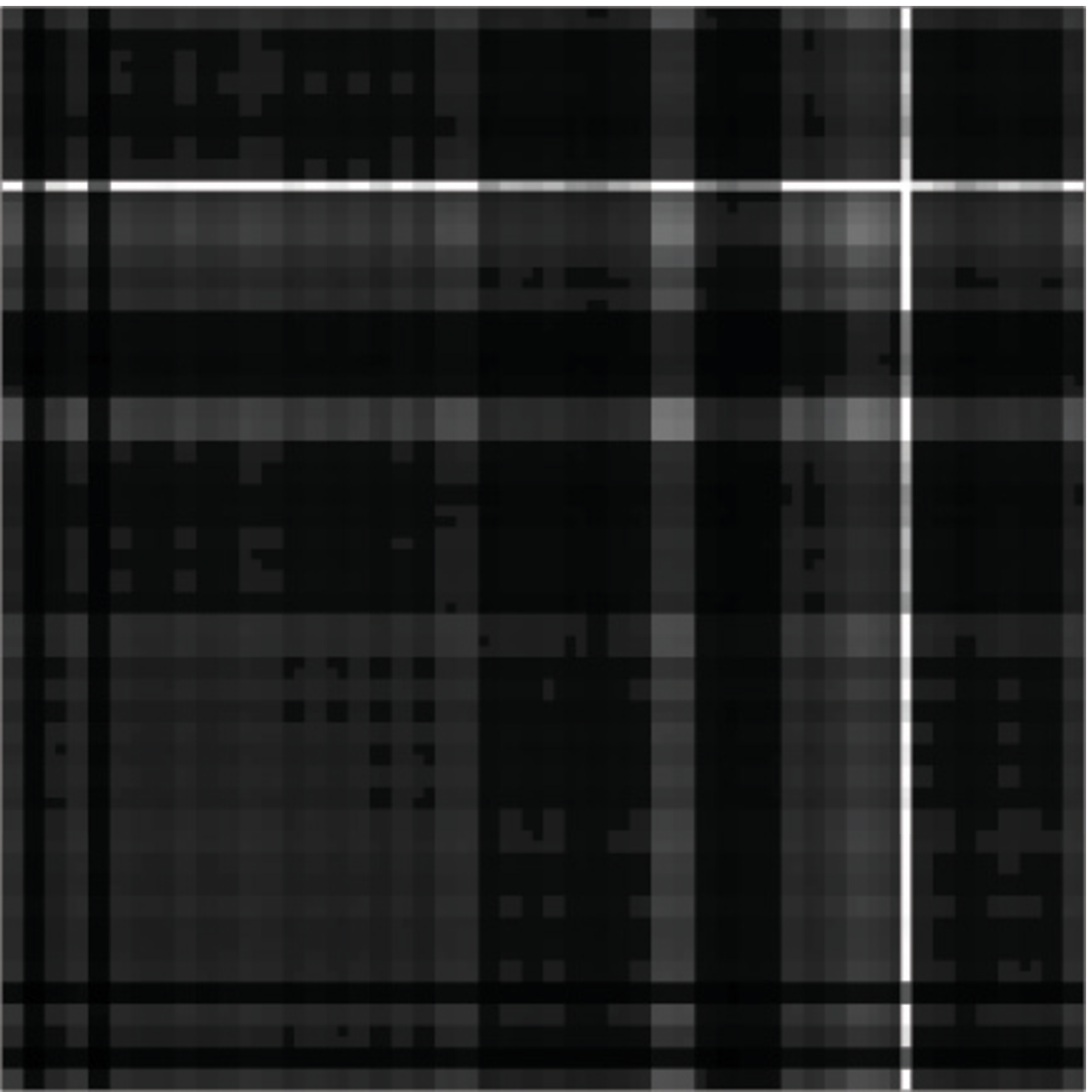}
      \caption{
  	LOCI trial: Flux and the autocorrelation matrix.
	Left: Case 1: Autocorrelation matrix $[ I_i ^k   I_i ^j ]$ for original reference library.
	Right: Case 2: Autocorrelation matrix for reference library with flux of image \#83 increased ten-fold.
	$x$ and $y$ axes are the images $I^j$ and $I^k$, color is degree of correlation.
	Color tables are stretched differently to bring out structure.
	  } \label{fig:autocorrelation}
\end{figure}

Figure~\ref{fig:autocorrelation} shows the autocorrelation matrix $[ I_i ^k   I_i ^j ]$ for each reference library in this trial.  The correlation of each image $I^j$ on the $x$-axis with each image $I^k$ on the $y$-axis is indicated by brightness in the figure, with white being most correlated and black being least correlated.  The effect of brightening the flux of image \#83 can clearly be seen on the right for Case 2.

\begin{figure}[htbp]
\centering
      \includegraphics[width=0.6\linewidth]{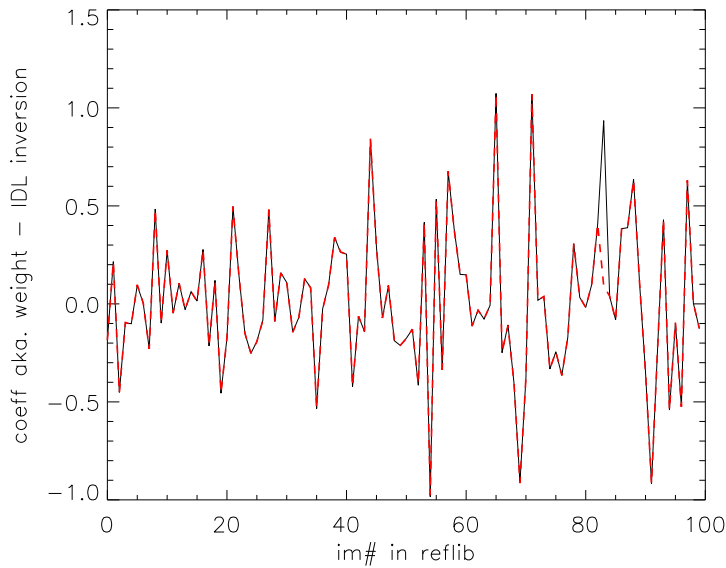}
      \caption{
  	LOCI-derived weights ($y$-axis) for each image in the library.
	Solid black line = Case 1: original reference library.
	Dashed red line = Case 2: reference library with flux of image \#83 increased ten-fold.
	LOCI automatically decreases the weight of image \#83 by a factor of ten in Case 2.
	  } \label{fig:lociweights}
\end{figure}

The LOCI-derived weights for both cases are shown in Fig.~\ref{fig:lociweights}.  The result of increasing the flux of image \#83 by a factor of ten is that LOCI decreases the weight for image \#83 by a factor of ten.  Hence we see that, although the autocorrelation matrix records higher values for correlated flux as well as correlated structure, LOCI independently selects for structure not flux correlation.  Therefore, it is not necessary to normalize the flux of images in the reference library.

\subsection{Snapshot LOCI processing}
Observations from epochs 1 and 2 result in approximately 1000 Keck images of 84 targets that we process with snapshot LOCI.  The LOCI algorithm is written in the IDL language and localized regions are chosen as annuli about the central star, of 5--16 pixels width.  A range of annulus widths are tried to test both the effect of a larger region and to test whether speckle location with respect to the annulus boundaries is important.  The annulus width and boundary locations are not found to have any significant effect on speckle subtraction, so we choose a default width of 16 pixels to optimize algorithm speed.

The algorithm minimizes the pixel-to-pixel noise within each local region; therefore, it is not necessary for the annular region to be processed in the shape of a 2-dimensional annulus.  Instead, each annulus is drawn out into a vector (as in Fig.~\ref{fig:reflib}) giving a 1-dimensional locality and a 2-dimensional autocorrelation matrix.  This improves processing speed greatly over a previous effort in which local regions were retained as 2-d images.  

The following image files are generated and written for each LOCI-processed image: the LOCI-derived PSF, the residuals, the residuals convolved with a FWHM-sized disk, the SNR map of the residuals, and the SNR map convolved with a FWHM-sized disk.  The SNR-map images are generated because they result in a flatter color table, the better for identifying point sources; similarly, the convolved images are generated to help visual identification of FWHM-sized point sources.  In cases with a bright wide secondary, LOCI is run twice on each image; once with the primary being subtracted and once with the secondary being subtracted to search for faint companions to either star.  We call this off-axis LOCI and discuss its efficacy in \S\ref{sec:offaxisloci}.

\subsection{Off-axis LOCI} \label{sec:offaxisloci}
\begin{figure}[htbp]
	\centering
      \includegraphics[width=0.75\linewidth]{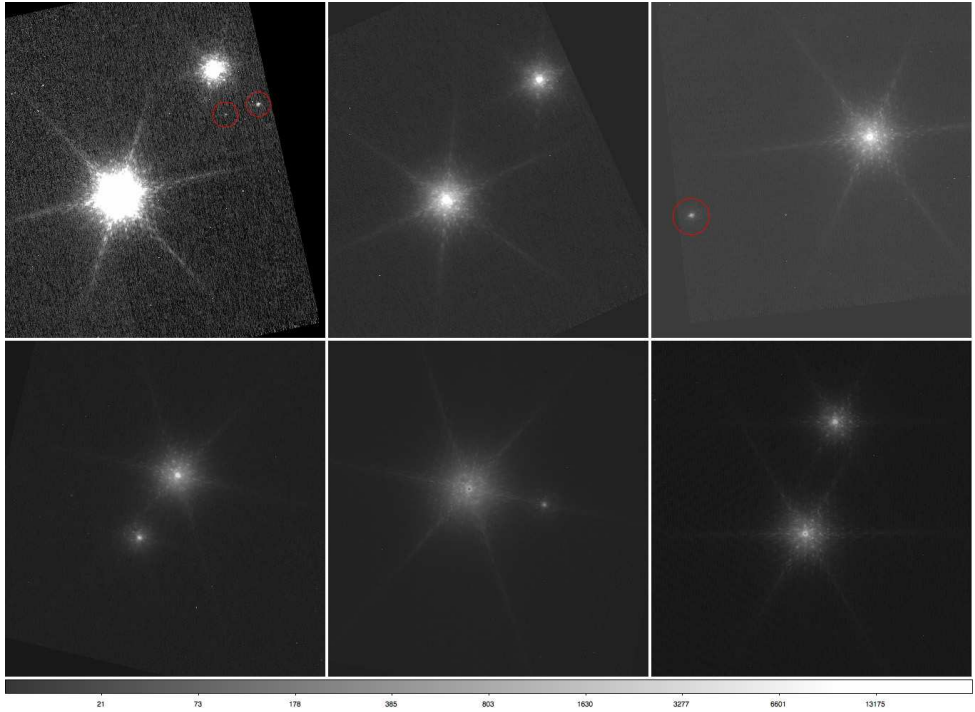}
      \caption{
	Sample of wide binaries suitable for off-axis LOCI on the secondary star (frames are $\sim$7'' across).
	  } \label{fig:widebinaries}
\end{figure}

The primary stars are all imaged such that their images are located on the same pixel near center of the NIRC2 detector; these are thus the on-axis stars with speckles caused by on-axis aberrations.  Eight targets of the 84 have a bright wide companion with its own speckles (examples shown in Fig.~\ref{fig:widebinaries}); these are the off-axis stars with speckles caused by off-axis aberrations.  Here we demonstrate use of the on-axis PSF to subtract off-axis speckles.

\begin{figure}[htbp]
	\centering
      \includegraphics[width=\linewidth]{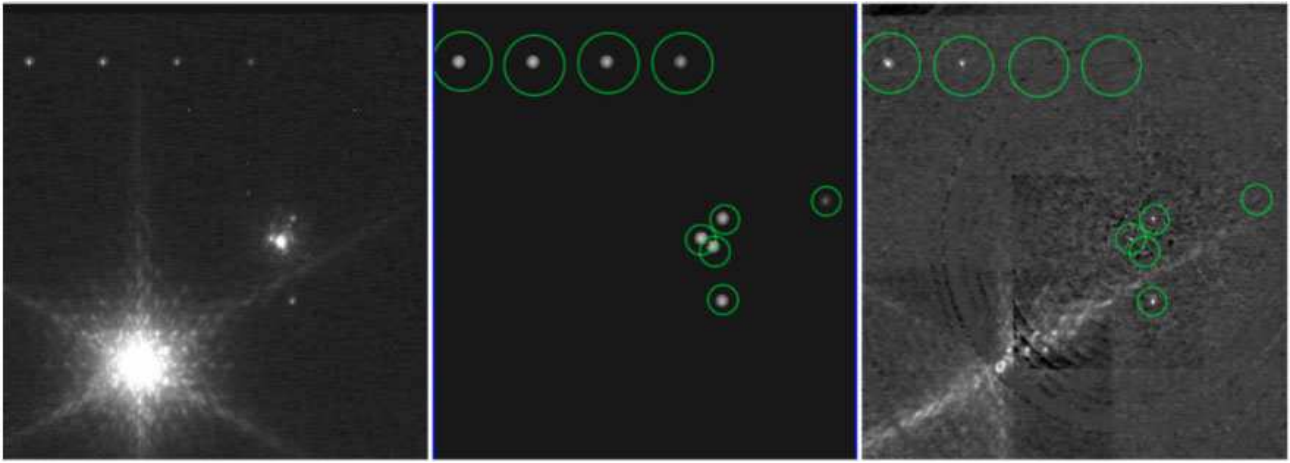}
      \caption{
	Demonstration of off-axis LOCI.
	Left: Original star plus 9 simulated brown dwarfs (center) added.
	Right: SNR map of residuals after running LOCI on the natural secondary.  Five of the nine simulated brown dwarfs are detected.
	  } \label{fig:testims}
\end{figure}

We demonstrate snapshot and off-axis LOCI here, using one of our targets that has a bright candidate companion $\sim$2\farcs5 to the northwest (see Fig.~\ref{fig:testims}, left).  Simulated brown dwarfs are added to these images, centered on the secondary star, and are processed to test for detectability and flux recovery.  The simulated brown dwarfs are created by taking the unsaturated core of a short-exposure image, scaling to various flux levels, and adding to the original image at strategic locations.

Figure~\ref{fig:testims} (right) illustrates detection of five out of nine simulated brown dwarfs about an off-axis secondary star.  Although the primary star retains a high number of bright speckles after off-axis LOCI, this method of using the on-axis reference library to subtract the off-axis PSF is indeed successful in revealing faint companions to bright secondary stars.  The contrast we achieve is discussed in \S\ref{sec:contrastcurves}.

\subsection{Contrast achieved} \label{sec:contrastcurves}
\begin{figure}[htbp]
	\centering
      \includegraphics[width=0.7\linewidth]{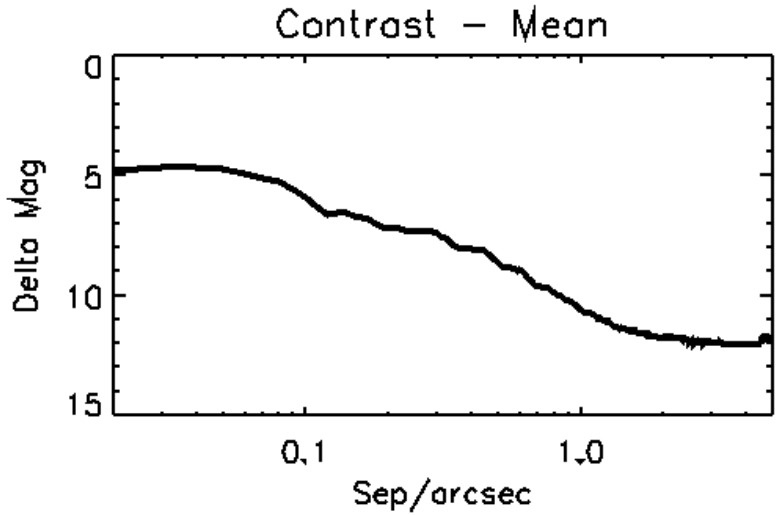}
      \caption{
      	Mean 5-$\sigma$ contrast for snapshot LOCI.
	} \label{fig:meancontrast}
\end{figure}

Contrast curves are generated from each LOCI-residual SNR map to determine detection limits.  The SNR map shows the signal-to-noise at each point in the image.  To generate this, we take the LOCI residual image and convolve it with a FWHM-sized disk, not normalizing the kernel.  This sums the flux within each photometry aperture.  Then, the standard deviation is determined within a 1-pixel-wide annulus, at each radial separation.  The flux in each pixel is divided by the standard deviation at that distance from the center.  The flux units are calibrated by normalization to the maximum flux measured in the unsaturated image and scaled by the integration time for the saturated image.  The result is then multiplied by 5 to give the SNR=5 detection floor.  Figure~\ref{fig:meancontrast} shows the mean 5-$\sigma$ contrast achieved in the survey (predominantly on-axis).

\begin{figure}[htbp]
	\centering
      \includegraphics[width=0.6\linewidth]{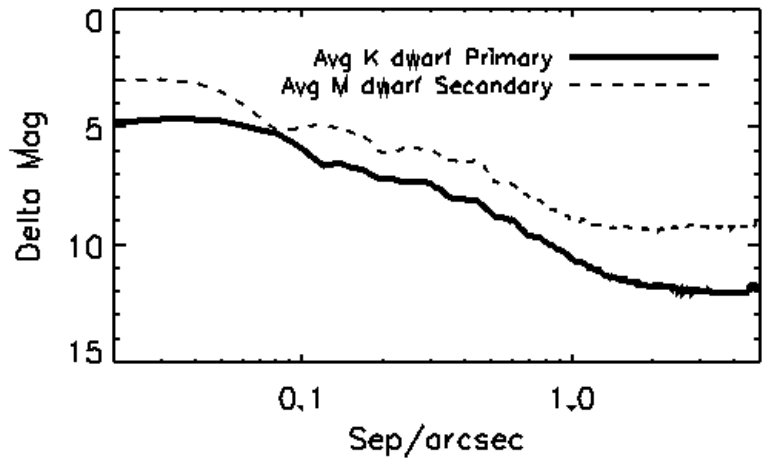}
      \caption{
      	Mean 5-$\sigma$ contrast: on-axis (solid line) and off-axis (dashed line) LOCI.
	} \label{fig:offaxiscontrast}
\end{figure}

Let us now compare the contrast achieved with off-axis LOCI to the mean contrast on-axis.  Figure~\ref{fig:offaxiscontrast} shows the off-axis contrast for the sample of bright wide secondaries.  It is lower than for on-axis LOCI, simply due to the secondary stars being (by definition) fainter than the primaries and so having a smaller normalization, but also due in some part to off-axis aberrations.

\begin{figure}[htbp]
	\centering
      \includegraphics[width=0.6\linewidth]{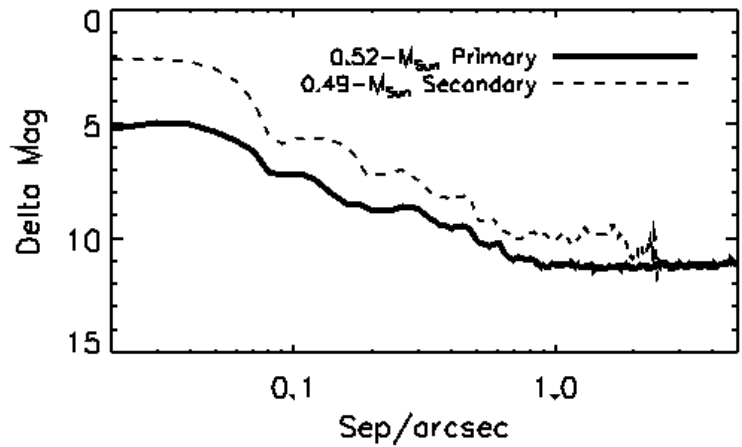}
      \caption{
      	Contrast: On- and off-axis LOCI for two stars of similar mass and brightness.
	} \label{fig:offaxiscontrast2}
\end{figure}

In order to disentangle whether the lower contrast for off-axis LOCI is due to lower normalization by a fainter star, or due to off-axis aberrations that produce different speckles than on-axis, let us compare the contrast for two stars of similar brightness.  One is the faintest primary star and the other is the brightest secondary star, and they have similar masses and magnitudes.  This is shown in Fig.~\ref{fig:offaxiscontrast2}, where we see that the on-axis contrast is indeed slightly better.  However, in \S\ref{sec:masslimits} we see that the limiting mass that can be detected is better on average for off-axis LOCI, simply because the secondary stars are fainter.

\section{Analysis and preliminary results}
Twenty-nine multiple systems are found and confirmed in the second epoch, including one brown dwarf and a handful of candidates still awaiting follow up.  Results here are preliminary until the brown dwarf(s) are spectroscopically confirmed (in prep).  This survey visually resolved 36 companions in 29 systems: 23 doubles, 5 triples, and 1 quadruple.

\subsection{Detection limits} \label{sec:masslimits}
Observations of apparent magnitude and distance, primarily from 2MASS\cite{skrutskie2006} and \textit{Hipparcos}\cite{hipparcos,hipparcos_catalog}, are used to transform the observed flux ratio and separation into mass and projected semimajor axis.  Recall from Fig.~\ref{fig:offaxiscontrast} that a brighter primary has better contrast for given LOCI performance because contrast is normalized to the maximum flux of the unsaturated star.  For this reason, the quantity of interest for the detection limit is mass rather than contrast.  Here we look at detectable mass limits.

\begin{figure}[htbp]
	\centering
      \includegraphics[width=0.6\linewidth]{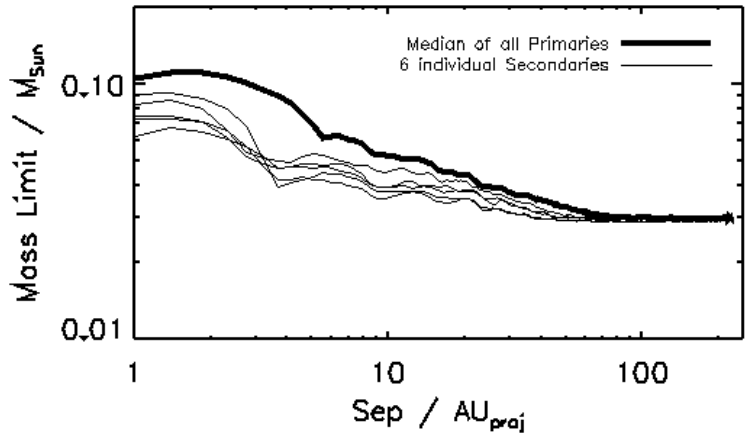}
      \caption{
      	Limiting mass detectable.
	On-axis (solid line) and off-axis (dashed line) LOCI, mean.
	} \label{fig:offaxismasslim}
\end{figure}

Each LOCI-residual image is used to generate a contrast curve for SNR=5.  Each contrast curve is then transformed from observable detection limits (delta-magnitude at each angular separation in arcseconds) to physical detection limits (companion mass at each projected semimajor axis) using 2MASS and \textit{Hipparcos} observables.  Thus we see in Fig.~\ref{fig:offaxismasslim} that the limiting detectable mass is improved for off-axis LOCI simply because the off-axis secondary stars are fainter than the on-axis primaries.

\begin{figure}[htbp]
\centering
      \includegraphics[height=0.7\linewidth,angle=90]{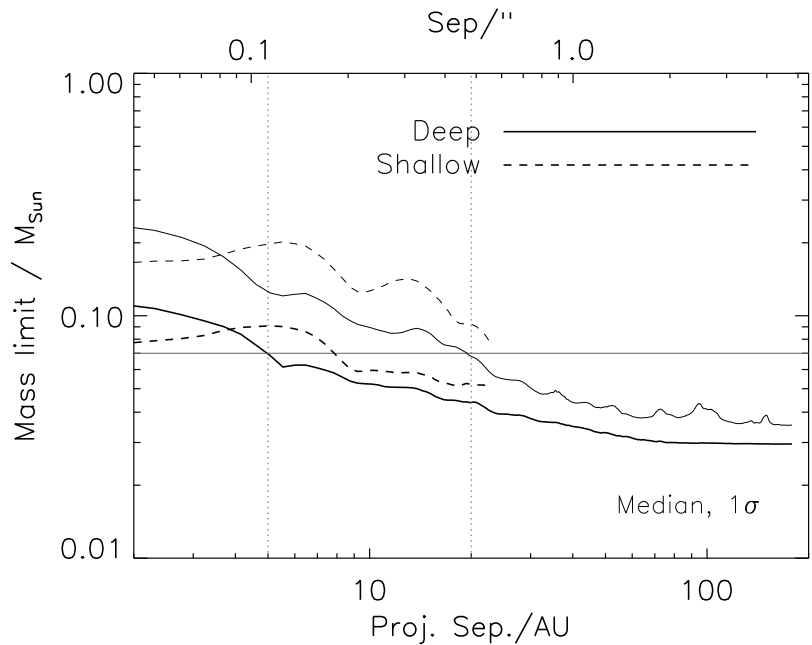}
	  \caption[Detection limits and completeness]{
	Detection limits and completeness.
	Median (thick lines) and median+1$\sigma$ (thin lines) curves for shallow (dashed lines) and deep exposures (solid lines).
	The horizontal line delineates the stellar/substellar boundary at 0.07 M$_{\sun}$.
	The stellar/substellar boundary coincides with the median detection curve at 5 AU and with the 1-$\sigma$ detection curve at 20 AU.
	Therefore, our survey as a whole is 1-$\sigma$ complete to SNR=5 brown dwarfs at separations beyond 20 projected AU.
	  } \label{fig:stats}
\end{figure}

Figure~\ref{fig:stats} shows the average limiting mass detectable as a function of projected separation.  The minimum detectable mass is 0.03 M$_{\sun}$, on average, near the L/T transition.  The unsaturated images (shallow exposures) give better contrast at the closest separations, where the saturated images (deep exposures) are affected at close separations by brighter speckles and a saturated core.
Figure~\ref{fig:stats} is marked to show the stellar/substellar boundary coincides with 5 projected AU for the median detection limit,
and 20 AU for 1$\sigma$.
We take the 1-$\sigma$ detection limit and can therefore say that our survey is complete to SNR=5 brown dwarfs orbiting beyond 20 projected AU.

\subsection{Multiplicity}
\begin{figure}[htbp]
\centering
      \includegraphics[width=0.4\linewidth]{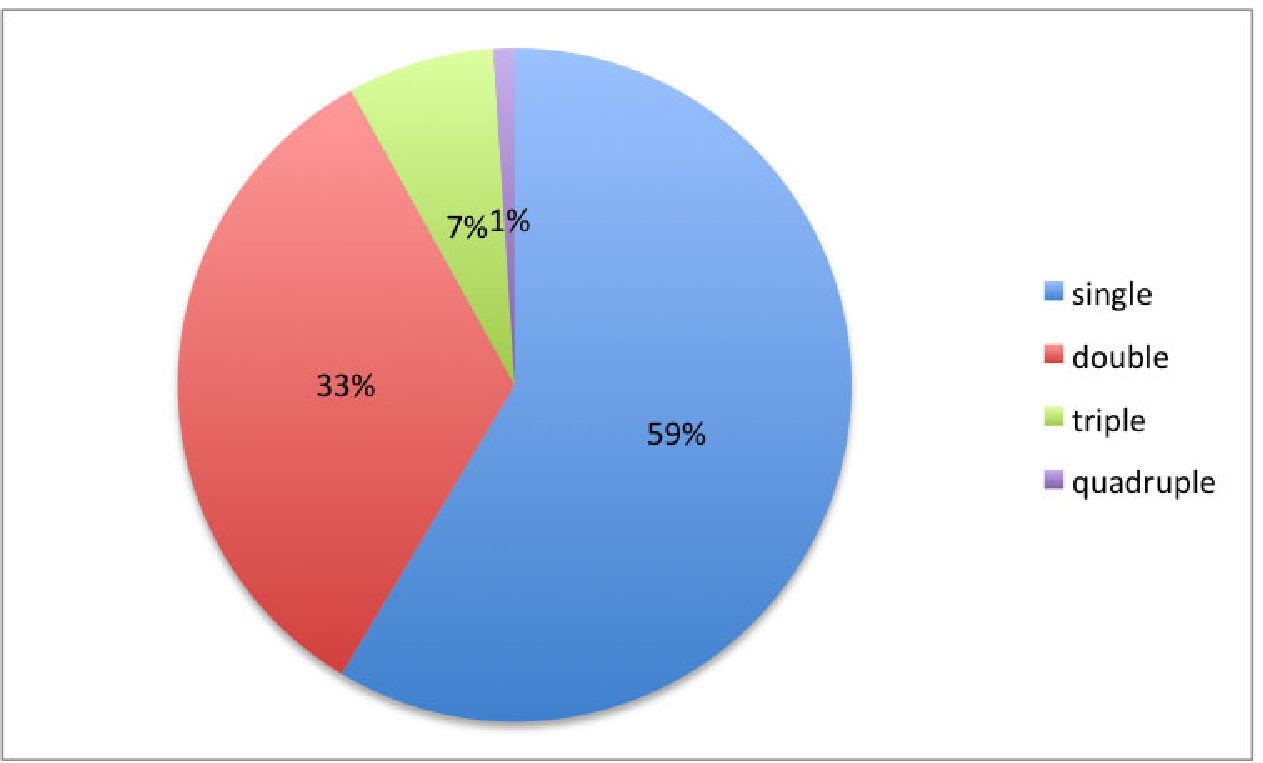}
      \includegraphics[width=0.4\linewidth]{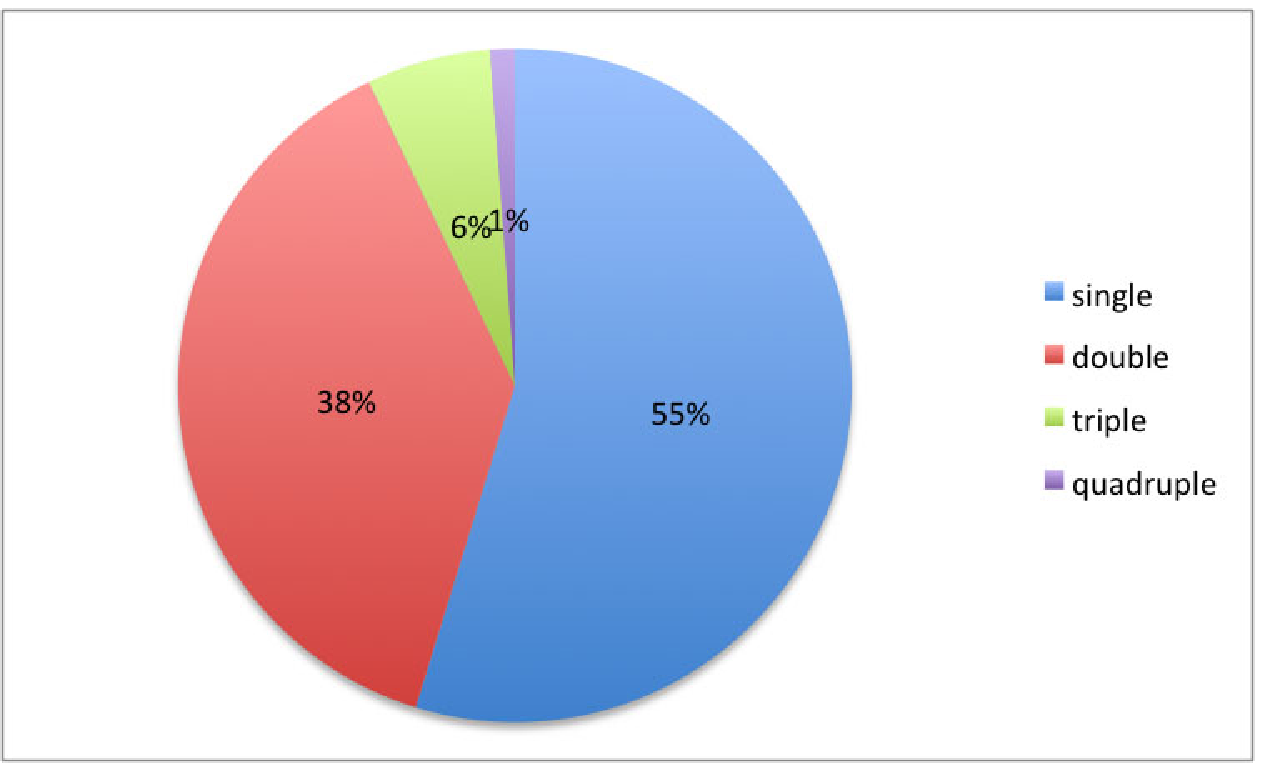}
	  \caption[Multiplicity]{
	  Frequency of multiple systems.
	  Left: Fischer \& Marcy\cite{fischer_marcy}, M-dwarf primaries (disregard the rounding error).
	  Right: This work, lower limit, including 10 speckle and spectroscopic binaries that remain unresolved in our images.
	  Half of our Hyades primaries are K dwarfs, and the remaining fraction is largely G and M dwarfs.
 	  } \label{fig:multiplicity}
\end{figure}

Fischer \& Marcy\cite{fischer_marcy} find 42\% of nearby M dwarfs host companions, taking radial velocity, speckle interferometry, and imaging surveys with 0.04--10$^4$ AU separations, and correcting for incompleteness.
Figure~\ref{fig:multiplicity}, left, shows the frequency of multiple systems in the Fischer survey, by number of components per system.

In this survey, 45\% of our Hyads are known to host multiple systems, out to $\sim$250 AU projected separation.  Figure~\ref{fig:multiplicity}, right, gives the multiplicity for our Hyades sample, including previously-known systems that are not resolved in this survey.  Ten targets that remain unresolved in this work are known to be spectroscopic or speckle binaries.

\section{Conclusions}
In this survey we implement snapshot LOCI to search GKM Hyads for low-mass companions, down to brown dwarfs at the L/T transition.
We find that the stellar multiplicity in the Hyades is at least $\sim$45\% including 1 quad, 5 triples, and 23 binaries, complete to brown dwarfs beyond 20 AU projected separation.  We have at least one new brown dwarf awaiting spectroscopic characterization.

LOCI improves the sensitivity by 1--4 magnitudes at varying separations.  With a rapid-fire observing strategy and no complicating coronagraphs nor ADI, the LOCI technique as employed in this survey results in contrasts of $10^{-2}$ at 0\farcs01, $10^{-4}$ at $<$1'', and $10^{-5}$ at 5''.
Using snapshot LOCI with a reference library consisting of a number of different stars imaged over several different nights, it is not necessary to normalize the flux; saturated and unsaturated PSFs can be used together in one reference library.

Depending on the imaging system in use, if most aberrations are in the pupil plane then off-axis LOCI could work as well as on-axis LOCI.  We find in the Keck case that off-axis LOCI works quite well to search for companions to bright wide binary stars, observed in position-angle mode: while off-axis LOCI resulted in lower contrasts, it resulted in similar or improved mass detection limits because the secondary stars are fainter.  Therefore, snapshot LOCI can find the best PSF for both primary and secondary stars---the latter being impossible with typical coronagraphs or ADI.  The snapshot LOCI technique used here can be implemented to search for faint companions among any set of well-correlated quasistatic-speckle-dominated PSFs.

\section*{ACKNOWLEDGMENTS}
This work was performed in part under contract with the Jet Propulsion Laboratory and is funded by NASA through the Sagan Fellowship Program under Prime Contract No.\ NAS7-03001.  JPL is managed for the National Aeronautics Space Administration (NASA) by the California Institute of Technology.
Any opinions, findings, and conclusions or recommendations expressed in this publication are those of the authors and do not necessarily reflect the views of the National Aeronautics Space Administration (NASA) or of The California Institute of Technology.
This research was supported in part by the University of California and National Science Foundation Science and Technology Center for Adaptive Optics, managed by the UC Santa Cruz under cooperative agreement No.~AST-9876783.
Portions of this work were performed under the auspices of the U.~S.~Department of Energy by the University of California, Lawrence Livermore National Laboratory under Contract W-7405-ENG-48.

\bibliography{ktmorz}
\bibliographystyle{spiebib}

\end{document}